# Discrete-Time I&I Adaptive Interconnection and Damping Passivity-Based Control for Nonlinearly Parameterized Port-Controlled Hamiltonian Systems

Mohammed Alkrunz[a,] *, Yaprak Yalçın[b]

[a]*Electrical and Electronics Engineering, Istanbul Aydin University, Istanbul, Turkey*
[b]*Control and Automation Engineering, Istanbul Technical University, Istanbul, Turkey*
*e-mail: mohammedalkrunz@aydin.edu.tr

**Abstract**—In this paper, a discrete-time I&I-based adaptive IDA-PBC controller for uncertain nonlinearly parameterized port-controlled Hamiltonian systems (PCH), where the parameter uncertainties are assumed in the energy function, is constructed. A proper formulation for the uncertain system dynamics is established where the uncertainties appear in nonlinearly parameterized form in the gradient of the Hamiltonian function. The adaptive IDA-PBC controller is constructed considering this formulation. For the adaptation mechanism of the IDA-PBC controller, a discrete-time parameter estimator is derived based on the immersion and invariance (I&I) approach. A structure for a free design function in the I&I-based estimator is proposed including some other free design functions. If these free design functions are selected to satisfy some conditions, derived in this paper, the Lyapunov asymptotic stability of the estimator dynamics is guaranteed. Besides, assuming these conditions are satisfied, local asymptotic stability of the closed-loop system, in a sufficiently large set is shown. The proposed method is applied to the two physical system examples and the performance of the adaptive controller is tested by simulation. It is demonstrated that the performance of the certain IDA-PBC controller is maintained by the adaptive IDA-PBC controller successfully.

## 1. INTRODUCTION

Recently, port-controlled Hamiltonian systems (PCH) have attracted more attention in non-linear theory and many works have been performed on PCH systems [1-8]. Hamiltonian system formulation has different considerable advantages since it reveals physical properties explicitly, such as being energy conservative or dissipative, which provides easy analysis [5]. The energy-shaped Hamiltonian function captures the total energy where this total energy consists of the potential and kinetic energies of the system and thus Hamiltonian function can be used as a candidate for the Lyapunov function [6-10]. By considering these advantages which are provided by the nice structure of PCH systems, PCH systems have utilized noticeable steps in the design of effective control law and they have been used in a wide range of practical control problems [11-17].

It is well known from control theories that uncertainties badly affect the control of real physical systems. These physical systems are sensitive to random uncertainties such as environment changes, parameter permutations, un-modeled dynamics, and measurement noises, and thus the formulation of these uncertainties into system models plays a big role in system modeling. It is difficult to achieve an accurate mathematical model that successfully captures the system dynamics. Therefore, it is very important to establish a control rule that can deal with these uncertainties and compensate for the corresponding resulting modeling errors and faults to improve the performance of the controlled system and stabilize the system at the desired operating points [6,16,14].

Adaptive control theory is a powerful control method for systems that are subject to parametric uncertainties. Generally, the designers try to automatically update the parameters of the constructed controller under some reasonable assumptions by providing a proper adaptation mechanism that compensates for uncertainties and parameter changes [17-20]. Adaptive control of uncertain PCH systems has been studied in several works in the literature [21-26].

Important interests and efforts are addressed day by day to achieve high-performance controllers for the systems which are modeled in Hamiltonian structure, especially for multivariable electromechanical systems [27]. The stability analysis of the nonlinear systems modeled as the port-controlled Hamiltonian systems is considered one of the most important research issues.

The Interconnection and Damping Assignment Passivity Based Control (IDA-PBC) is considered as a powerful stabilization control design method which is developed by [28] concerning passivation. In other words, in this methodology, a storage function that has its minimum at the desired equilibrium points is used by the stabilizing controller to passivize the given actual system [29-34]. Thus, the closed-loop energy function in the IDA-PBC approach is determined by solving the partial differential equations (PDEs) that are obtained by selecting the desired energy function in addition to the interconnection and damping structure matrices [28]. Unfortunately, it is a challenging and not easy issue to obtain solutions to these PDEs for underactuated systems.

On the other hand, Astolfi and Ortega in [35] have presented a new tool for nonlinear stabilizing control and adaptive control. They used two classical tools of geometric nonlinear control and nonlinear regulator theory for general nonlinear systems, namely, system immersion and manifold invariance, to reduce the design problem of the stabilization and adaptive control rules to other sub-problems, which are easier to solve. This method is called the Immersion and Invariance (I&I) approach. More precisely, the I&I stabilization and the I&I adaptive control. Some studies use the I&I approach as a stabilizing controller [36-38] and some other studies use it as an adaptive controller by establishing I&I parameter estimation [39-52].

To the best of our knowledge, and out of our previous study [21], where the linearly parameterized case is considered, there are only three studies on adaptive IDA-PBC for linearly parameterized Hamiltonian systems in the literature [53-55]. More precisely, like our study, the authors assumed the existence of parametric uncertainties in the system model, and unlike our study, the authors in [53-54] expressed the uncertainties as uncertain control input and thus the adaptive controller compensates for these errors while the authors in [55] presented two adaptive control approaches to handle uncertainties caused by parametric and modeling errors. Namely, adaptive IDA-PBC controller for friction estimation and for potential energy function estimation where the authors considered the gradient of the potential energy function can be represented as a linearly parameterized expression. Besides, all the mentioned studies [53-55] proposed controllers in continuous-time settings considering linearly parameterized cases while our study proposed the controller in a discrete-time setting and the nonlinearly parameterized case is considered.

Indeed, parametric uncertainties cannot be always formulated in a linearly parameterized form, and in the literature, we didn't encounter a study that considers this issue for PCH systems. This paper presents a discrete-time adaptive IDA-PBC control based on the I&I approach for the non-linearly parameterized port-controlled Hamiltonian system (PCH) with parameter uncertainties in the Hamiltonian (energy) function. Namely, the uncertainties are assumed in a non-linearly parameterized form in the gradient of the Hamiltonian function.

This paper combines the IDA-PBC technique with the I&I approach in the discrete-time setting to establish an adaptive control method. Namely, we constructed the IDA-PBC controller for the uncertain non-linearly parameterized PCH systems where the parameters of the IDA-PBC controller are tuned by the I&I-based parameters estimator. Firstly, the stability conditions for parameter estimation error dynamics are established by referring to the P-monotonicity and being locally Lipschitz continuous concepts. A structure that facilitates the design of an estimator that satisfies the derived stability condition is proposed for the free design function of the I&I estimator. Note that, the proposed structure has some other free design function also to enable to assign of the desired dynamics to the estimator error dynamics, and thus the stability of the estimator error system is ensured. Afterward, the local asymptotic stability of the closed-loop system, which is obtained with the discrete-time adaptive IDA-PBC controller that uses the estimated parameters provided by the I&I-based estimator, is shown in any sufficiently large set by assuming the same conditions that are derived for the stability of the estimator. The proposed adaptive control method is employed for fully-actuated and under-actuated physical systems and the construction of the free function in the proposed structure is given for each example. The simulation results elaborated that the constructed asymptotically stable estimators successfully estimate the uncertain parameters, and then eventually the closed-loop system tracks the desired closed-loop system which would be obtained by the IDA-PBC controller with true parameters. Namely, eventually, the desired Hamiltonian function is successfully assigned to the proposed uncertain port-controlled Hamiltonian function, where all the trajectories of the closed-loop system are bounded and converge to the defined equilibrium points that are to say the entire closed-loop system is stabilized at the minimum of the desired energy function.

The remaining of the paper is structured as follows. In section 2, we provide some preliminaries on the discrete-time modeling of PCH and the I&I adaptive control problem formulation for general nonlinear systems and the corresponding conditions in the discrete-time setting. In section 3, firstly, the formulation of the considered adaptive control problem is presented, afterward, I&I parameter estimator construction is performed and the conditions on the stability of the estimator are derived, and finally proposed adaptive IDA-PBC controller is given together with the stability analysis of the closed-loop system. In section 4, two nonlinearly parameterized examples are considered to investigate the applicability and the performance of the proposed adaptive control method, and the results of the corresponding simulation are given. Finally, the conclusions are given in Section 5.

## 2. PRELIMINARIES

### 2.1 Modeling of PCH and IDA-PBC Formulation in Discrete-Time

Consider the port-controlled Hamiltonian system (PCH) with dissipation in the local coordinates in the continuous-time setting as follows:

$$\dot{x}(t) = [J(x) - R(x)]\, \nabla H(x) + g(x)\, u(t) \tag{2.1a}$$

$$\nabla H(x) := grad\, H(x) = \left[\frac{\partial H}{\partial x_1} \quad \cdots \quad \frac{\partial H}{\partial x_n}\right]^T \tag{2.1b}$$

$$x = [x_1, \quad \cdots \quad , x_n]^T \tag{2.1c}$$

where $x(t) \in \Re^n$ is the state vector, $u(t) \in \Re^m$ is the control input, $g(x) \in \Re^{n \times m}$ is the input force matrix, matrices $J(x) = -J^T(x)$ and $R(x) = R^T(x) \geq 0$ are the $n \times n$ standard skew-symmetric matrix and the non-negative dissipation structure matrix respectively where $J(x)$ determines the internal connection structure of the system while $R(x)$ represents the dissipation, and $H(x): R^n \to R$ is the Hamiltonian function of the system, namely, the total energy function of the system.

In this paper, we employ the method proposed in [56] to build a discrete-time model of port-controlled Hamiltonian systems. Namely, we obtained the discrete-time models by replacing the gradient terms in the continuous-time model with some discrete gradients that satisfy the conditions defined in [57] and by substituting forward-Euler approximations for the time derivatives in the continuous-time model. The mentioned discrete gradient conditions are given below.

**Definition 1.** Let $H(x)$ be a differential scalar function in $x \in \Re^n$, then $\bar{\nabla}H(x_k, x_{k+1})$ is a discrete gradient of $H$ if it is continuous in $x$ and,

$$\bar{\nabla}H^T(x_k, x_{k+1})[x_{k+1} - x_k] = H(x_{k+1}) - H(x_k) \tag{2.2a}$$

$$\bar{\nabla}H^T(x_k, x_k) = \bar{\nabla}H^T(x_k). \tag{2.2b}$$

Using forward-Euler approximation for the derivatives of state variables in (2.1) as below:

$$\dot{x} \cong \frac{x_{k+1} - x_k}{T} \tag{2.3}$$

and replacing the gradient term $\nabla H(x)$ with the discrete gradient $\bar{\nabla}H(x_k, x_{k+1})$ yields the following discrete-time model of the port controlled Hamiltonian systems [56]:

$$x_{k+1} - x_k = T\, [J(x_k) - R(x_k)]\, \bar{\nabla}H(x_k) + Tg(x_k)\, u_k \tag{2.4a}$$

$$y_k = g^T(x_k)\bar{\nabla}H(x_k) \tag{2.4b}$$

where $\bar{\nabla}H$ is any discrete gradient that meets the conditions stated in Definition 1 and $T$ is the sampling time. In the examples that we consider to show the applicability of the adaptive controller proposed in this study, the following discrete gradient provided in [56] is used considering $\nabla H(x)$ can be formulated as $\nabla H(x) = Q(x)x$;

$$\bar{\nabla}H(x_k, x_{k+1}) = \Phi_k[x_{k+1} + x_k] \tag{2.5a}$$

$$\Phi_k = \frac{1}{2} Q(x) \bigg|_{x = \frac{x_{k+1}+x_k}{2}} \qquad (2.5b)$$

Here, by approximating $x_{k+1}$ with $x_{k+1} = 2x_k - x_{k-1}$, the discrete gradient can be directly obtained as given below:

$$\bar{\nabla} H(x_k, x_{k+1}) = \nabla H(x) \bigg|_{x = \frac{3x_k - x_{k-1}}{2}} \qquad (2.6)$$

In the work of [56], a discrete-time IDA-PBC design is given in generalized coordinates where it is assumed that the continuous-time model of the desired Hamiltonian system is available. Following the same assumption in [56], the discrete-time IDA-PBC formulation in local coordinates can be established as we did in our previous study in [21] and restated below.

Consider the following desired discrete-time Hamiltonian system:

$$x_{k+1}^d - x_k^d = T [J_d(x_k) - R_d(x_k)] \bar{\nabla} H_d(x_k) \qquad (2.7)$$

where $J_d(x_k) = -J_d^T(x_k)$ and $R_d(x_k) = R_d^T(x_k)$. In this case, $H_d$ and $J_d$ should be constructed such that the following matching condition holds [30]:

$$g^\perp \{J_d(x_k) \bar{\nabla} H_d(x_k) - [J(x_k) - R(x_k)]\bar{\nabla} H(x_k)\} = 0 \qquad (2.8)$$

with $g^\perp$ a full rank left annihilator of g, i.e., $g^\perp g = 0$. Note that in the case of fully actuated Hamiltonian systems the above condition is satisfied for any desired system. The IDA-PBC controller is composed of an energy shaping control rule and a damping injection control rule. Assume $J_d(x_k)$ and $H_d(x_k)$ could be found satisfying the condition (2.8), then the energy shaping control rule is derived as:

$$u_{es} = [g^T(x_k)g(x_k)]^{-1}g^T(x_k)\{J_d(x_k) \bar{\nabla} H_d(x_k) - [J(x_k) - R(x_k)]\bar{\nabla} H(x_k)\}. \qquad (2.9a)$$

Because the closed-loop system obtained with this control rule is also a Hamiltonian system with the desired energy function, the damping injection control rule to make the closed-loop system asymptotically stable can be formed as follows:

$$u_{di} = -K_v \, g^T(x_k) \, \bar{\nabla} H_d(x_k). \qquad (2.9b)$$

As a result, the entire discrete-time IDA-PBC control rule is established by adding these two control rules together, as shown below:

$$u_k = u_{es} + u_{di}. \qquad (2.9c)$$

## 2.2 Immersion and Invariance Control Approach in Discrete-time

Astolfi et al. [35] are the first to present the I&I methodology in the continuous-time domain. Following the approach, the main premise of utilizing the I&I technique to achieve stabilization is to establish an asymptotically stable target system with an order strictly lower than the dimension of the considered system and to construct a control rule that immerses the system dynamics in the selected target system. Katakura et al. [58] proposed the discrete-time I&I formulation, which has been used in numerous researches [38-39]. The following is the list of the corresponding I&I conditions.

Assume the nonlinear discrete-time system is given as follows:

$$x_{k+1} = f(x_k) + g(x_k)u_k \tag{2.10}$$

where $x_k \in \Re^n$ denotes the system's states and $u_k \in \Re^m$ denotes the control vector, and $x^* \in \Re^n$ denotes the equilibrium point to be stabilized. If the nonlinear system in (2.10) meets the following four conditions (H1 to H4), the equilibrium point $x^*$ is a (globally) asymptotically stable equilibrium of the closed-loop system:

$$x_{k+1} = f(x_k) + g(x_k)\psi(x_k, \phi(x_k)) \tag{2.11}$$

with mappings $\phi(.): \Re^n \to \Re^{n-p}$ and $\psi(.,.): \Re^{n \times (n-p)} \to \Re^m$ and $p < n$.

**H1**: (Target *Dynamics*): The target system,

$$\xi_{k+1} = \alpha(\xi_k) \tag{2.12}$$

with state $\xi \in \Re^p$ and mapping $\alpha(.): \Re^p \to \Re^p$ has a globally asymptotically stable equilibrium at $\xi^* \in \Re^p$ and $x^* = \pi(\xi^*)$ with mapping $\pi(.): \Re^p \to \Re^n$.

**H2**: (*Immersion and Invariance Condition*): For all $\xi \in \Re^p$,

$$f(\pi(\xi_k)) + g(\pi(\xi_k))\, c(\pi(\xi_k)) = \pi(\alpha(\xi_k)) \tag{2.13}$$

with mapping $c(.): \Re^p \to \Re^m$.

**H3**: (*Implicit Manifold*): The following identity between sets holds,

$$\{x_k \in \Re^n | \phi(x_k) = 0\} = \{x_k \in \Re^n | x_k = \pi(\xi_k)\} \tag{2.14}$$

**H4**: (*Manifold Attractively and Trajectory Boundedness*): All trajectories of the system,

$$z_k = \phi(x_k) \tag{2.15a}$$
$$x_{k+1} = f(x_k) + g(x_k)\,\psi(x_k, z_k) \tag{2.15b}$$

are bounded for all $k \geq 0$ and satisfy,

$$\lim_{k \to \infty} z_k = 0.$$

The discrete-time adaptive control problem formulation via immersion and invariance approach can be established as below similarly to the continuous-time formulation provided in [35].

Assume that there exists a parameterized function $\psi(x_k, \theta)$, where $\theta \in \Re^q$, such that the system in (10) is stabilized for some unknown $\theta^* \in \Re^q$ and the system in (2.16) has a globally asymptotically stable equilibrium point at $x_k = x_k^*$.

$$x_{k+1} = f^*(x_k) = f(x_k) + g(x_k)\psi(x_k, \theta^*) \tag{2.16}$$

Then, the system in (2.10) is said to be adaptively I&I stabilizable if the system

$$x_{k+1} = f(x_k) + g(x_k)\, \psi(x_k, \theta_k^{est}) \tag{2.17a}$$
$$\hat{\theta}_{k+1} = \alpha\, (x_k, \theta_k^{est}) \tag{2.17b}$$
$$\theta_k^{est} = l\, (x_k, \hat{\theta}_k) \tag{2.17c}$$

where $\alpha\, (x_k, \theta_k^{est})$ is the update law and $\theta_k^{est}$ is the parameter estimation, is I&I stabilizable with the target dynamics

$$\xi_{k+1} = f^*(\xi_k). \tag{2.18}$$

Therefore, the problem to be considered is to establish (2.17b) and (2.17c) together with $\psi(x_k, \theta^*)$ such that (2.10) is adaptively I&I stabilizable.

## 3. MAIN RESULTS

### 3.1 Problem Formulation

Consider the discrete-time model of the port-controlled Hamiltonian systems in (2.4) and assume that there are uncertainties in the energy function. The discrete-time model of the non-linearly parameterized uncertain port-controlled Hamiltonian system may be expressed as follows:

$$x_{k+1} - x_k = T[J(x_k) - R(x_k)] \left[\bar{\nabla}\widetilde{H}_{un}(x_k, \theta) + \bar{\nabla}H_{kn}(x_k)\right] + T\, g(x_k)\, u_k \tag{3.1}$$

where $\bar{\nabla}\widetilde{H}_{un}(x_k, \theta)$ is non-linearly parameterized as:

$$x_{k+1} - x_k = T[J(x_k) - R(x_k)]\, (\underbrace{\Pi(x_k)\, \emptyset(x_k, \theta)}_{\bar{\nabla}\widetilde{H}_{un}} + \bar{\nabla}H_{kn}(x_k)) + T\, g(x_k)\, u_k \tag{3.2}$$

and $x_k \in \Re^n$ is the state vector which is assumed measurable, $\bar{\nabla}H_{kn} \in \Re^n$ is the known part of the discrete gradient, $\bar{\nabla}\widetilde{H}_{un} \in \Re^n$ is the unknown part of the discrete gradient that contains the unknown parameters, $u_k \in \Re^m$ is the control vector, $g \in \Re^{n \times m}$ is the input force matrix, and $\theta = [\theta_1 \quad \theta_2 \quad \cdots \quad \theta_p]^T \in \Re^s$ a constant vector representing the unknown parameters. By considering that the system is non-linearly parameterized, the unknown part $\bar{\nabla}\widetilde{H}_{un}$ can be always formalized in the form of $\Pi(x_k)\emptyset(x_k, \theta)$, where all the known terms and system's states are included inside $\Pi(x_k) \in \Re^{n \times r}$, while the unknown terms together with the system's states are the variables of the function $\emptyset(x_k, \theta) \in \Re^{r \times 1}$.

Here, let re-write (3.2) as below:

$$x_{k+1} - x_k = T[J(x_k) - R(x_k)]\, \Pi(x_k)\, \emptyset(x_k, \theta) + T[J(x_k) - R(x_k)]\bar{\nabla}H_{kn}(x_k) + T\, g(x_k)\, u_k \tag{3.3}$$

Now we can state that the problem considered is in the sequel is constructing a discrete-time adaptive IDA-PBC controller $u_k$ that matches the considered uncertain system (3.3) in the closed-loop with the desired discrete-time Hamiltonian system given in (2.7) and thus stabilizes the closed-loop system at the minimum of the desired energy function successfully.

## 3.2 I&I Parameter Estimation Design

Here the design of the parameter estimator is performed based on the Immersion and Invariance (I&I) approach for the uncertain non-linearly parameterized Hamiltonian systems given in (3.3) and recalled below.

$$x_{k+1} = T[J(x_k) - R(x_k)]\Pi(x_k)\emptyset(x_k, \theta) + T[J(x_k) - R(x_k)]\bar{\nabla}H_{kn}(x_k) + Tg(x_k)u_k + x_k \quad (3.4)$$

Here, the design of a discrete-time adaptive state feedback control law of the following form is considered:

$$\hat{\theta}_{k+1} = \alpha\left(x_k, \theta_k^{est}\right) \quad (3.5a)$$
$$u_k = v\left(x_k, \theta_k^{est}\right) \quad (3.5b)$$

such that all trajectories of the closed-loop system are bounded and

$$\lim_{k\to\infty} x_k = x_k^*$$

where $x_k^*$ is the desired equilibrium point to be stabilized. Then, let us define $l(x_k, \hat{\theta}_k)$ as follows to solve this problem:

$$l(x_k, \hat{\theta}_k) = \hat{\theta}_k + \beta(x_{k-1})x_k \quad (3.6)$$

and the parameter estimation error is defined as:

$$z_k = \hat{\theta}_k + \beta(x_{k-1})x_k - \theta = \theta_{est} - \theta. \quad (3.7)$$

Hence,

$$z_{k+1} = \hat{\theta}_{k+1} + \beta(x_k)x_{k+1} - \theta. \quad (3.8)$$

Therefore, by subtracting (3.7) from (3.8), and by considering the system dynamics in (3.4), then

$$\begin{aligned} z_{k+1} - z_k &= \hat{\theta}_{k+1} - \hat{\theta}_k - \beta(x_{k-1})x_k + \beta(x_k)x_{k+1} \\ &= \hat{\theta}_{k+1} - \hat{\theta}_k - \beta(x_{k-1})x_k + \beta(x_k)[T(J-R)\bar{\nabla}H_{kn} + Tgu_k + x_k] \\ &\quad + \beta(x_k)T(J-R)\Pi(x_k)\emptyset(x_k, \theta) \end{aligned} \quad (3.9)$$

can be obtained. At this point, selecting the update law as follows:

$$\hat{\theta}_{k+1} = \hat{\theta}_k + \beta(x_{k-1})x_k - \beta(x_k)[T(J-R)\bar{\nabla}H_{kn} + Tgu_k + x_k] \\ - \beta(x_k)T(J-R)\Pi(x_k)\emptyset(x_k, \theta_k^{est}) \quad (3.10)$$

yields the parameter estimation error dynamics:

$$z_{k+1} - z_k = \beta(x_k)T(J-R)\Pi(x_k)[\emptyset(x_k, \theta) - \emptyset(x_k, \theta_k^{est})] \quad (3.11)$$

Here, the following definitions and assumptions are considered to establish sufficient conditions for the stability of the parameter estimation error dynamics in (3.11).

**Definition 2**. [59] Given a matrix $P \in \Re^{s \times s}$ and $P = P^T > 0$. A mapping $L: \Re^s \to \Re^s$ is P-monotone [resp. strictly P-monotone] if and only if, for all $a, b \in \Re^s$.

$$(a - b)^T P [L(a) - L(b)] \geq 0$$

$$[resp. \quad (a - b)^T P [L(a) - L(b)] > 0, \quad \forall a \neq b]$$

**Definition 3**. [60] Given a function $f: \Re^s \to \Re^s$ is called Locally Lipschitz at $x_o \in \Re^s$ if there exists a neighborhood $B(x_o, \varepsilon)$ with $\varepsilon > 0$ and $L > 0$ such that:

$$\|f(x_1) - f(x_2)\| \leq L\|x_1 - x_2\|$$

for all $x_1$ and $x_2 \in B(x_o, \varepsilon)$ such that $B(x_o, \varepsilon) := \{x \in \Re^s | \|x - x_o\| < \varepsilon\}$ where L is called the "Lipschitz Constant" and in general $L$ depends on $x_o$ and $\varepsilon$.

**Assumption 1**. Define the parameterized mapping $\psi(x, \theta): \Re^s \to \Re^s$:

$$\psi(x_k, \theta) := \beta(x_k) \underbrace{T(J - R) \Pi(x_k)}_{A(x_k) \in R^{n \times r}} \phi(x_k, \theta) = \beta(x_k) A(x_k) \phi(x_k, \theta) \tag{3.12a}$$

$$[resp. \quad \psi(x_k, \theta_k^{est}) := \beta(x_k) A(x_k) \phi(x_k, \theta_k^{est})] \tag{3.12b}$$

There exists a set $\Omega \subset \Re^n$ such as the mapping $\psi(x_k, \theta)$ is P-monotone [resp. strictly P-monotone] concerning $\theta$ for all $x_k \in \Omega$.

**Assumption 2**. Consider the parameterized mapping $\psi(x, \theta): \Re^s \to \Re^s$ in (3.12). Then, there exists a set $\Omega \subset \Re^n$ such as the mapping $\psi(x_k, \theta)$ satisfies (3.13) with $L \leq 1$. (Namely, $\psi(x_k, \theta)$ is Lipschitz for all $x_k \in \Omega$ concerning $\theta$ with $L \leq 1$).

$$\|\psi(x_k, \theta) - \psi(x_k, \theta_k^{est})\| \leq L\|\theta - \theta_k^{est}\| \tag{3.13}$$

**Proposition.** Consider the uncertain discrete-time non-linearly parameterized Hamiltonian system given in (3.3), parameter estimation errors in (3.7), and parameter update law (3.10). Let the free design function $\beta(x_k)$ is selected such that Assumption 1 and Assumption 2 are satisfied, namely $\psi(x_k, \theta)$ in (3.12) is P-monotone and Lipschitz with $L \leq 1$. Then, the estimation error dynamics (3.11) is Lyapunov stable (Lyapunov asymptotically stable if Assumption 1 and Assumption 2 are satisfied strictly).

**Proof.** Consider that the system equation given in (3.3) and the parameter estimation errors defined in (3.7) and (3.8) and that the expression in (3.11) is obtained for parameter estimation dynamics and the parameter update law is selected as given in (3.10). Now, let's choose a candidate Lyapunov function as follows:

$$V_{z_k} = z_k^T P z_k \tag{3.14}$$

and respectively $V_{z_{k+1}} = z_{k+1}^T P z_{k+1}$. Hence, the time difference of the function (3.14) is obtained as follows:

$$\Delta V_{z_k} = V_{z_{k+1}} - V_{z_k} = z_{k+1}^T P z_{k+1} - z_k^T P z_k \quad (3.15)$$

To prove the stability of the estimator, we need to show that (3.15) is negative semi-definite, i.e., $\Delta V_{z_k} \leq 0$ (negative definite for asymptotic stability, i.e., $\Delta V_{z_k} < 0$). Now, let us derive the conditions on the negative semi-definiteness (negative definiteness) of (3.15).

Here, notice that the $\Delta V_{z_k}$ can be written as the sum of the following two terms.

$$z_{k+1}^T P(z_{k+1} - z_k) = z_{k+1}^T P z_{k+1} - z_{k+1}^T P z_k \quad (3.16a)$$
$$z_k^T P(z_{k+1} - z_k) = z_k^T P z_{k+1} - z_k^T P z_k \quad (3.16b)$$

Therefore, by proving that both (3.16a) and (3.16b) are smaller than or equal to zero, then $\Delta V_{z_k} \leq 0$ (strictly smaller than zero for $\Delta V_{z_k} < 0$) will be proved.

Firstly, let us prove (3.16a), namely, $z_{k+1}^T P(z_{k+1} - z_k) \leq 0$. When each term in (3.16a) is replaced with its equivalent,

$$z_{k+1}^T P(z_{k+1} - z_k) = \left[\hat{\theta}_{k+1} + \beta(x_k) x_{k+1} - \theta\right]^T P \left[\beta(x_k) A(x_k)[\phi(x_k, \theta) - \phi(x_k, \theta_k^{est})]\right] \quad (3.17)$$

is obtained and replacing $\hat{\theta}_{k+1}$ with (3.10) and $x_{k+1}$ with (3.3) yields:

$$z_{k+1}^T P(z_{k+1} - z_k) \quad (3.18)$$
$$= \left[(\theta_k^{est} - \theta) + \beta(x_k) A(x_k)[\phi(x_k, \theta) - \phi(x_k, \theta_k^{est})]\right]^T P \left[\beta(x_k) A(x_k)[\phi(x_k, \theta) - \phi(x_k, \theta_k^{est})]\right]$$
$$= \underbrace{(\theta_k^{est} - \theta)^T P(\beta(x_k) A(x_k)[\phi(x_k, \theta) - \phi(x_k, \theta_k^{est})])}_{I}$$
$$+ \underbrace{\begin{array}{l}(\beta(x_k) A(x_k)[\phi(x_k, \theta) - \phi(x_k, \theta_k^{est})])^T P \times \\ (\beta(x_k) A(x_k)[\phi(x_k, \theta) - \phi(x_k, \theta_k^{est})])\end{array}}_{II}$$

Therefore, if $I \leq 0$ and $\|I\| \geq \|II\|$, then $z_{k+1}^T P(z_{k+1} - z_k) \leq 0$. Now, by referring to Definition 2 and Assumption 1, we can claim the following:

$$I = -\underbrace{(\theta - \theta_k^{est})^T P [\psi(x_k, \theta) - \psi(x_k, \theta_k^{est})]}_{\geq 0} \leq 0 \quad (3.19a)$$

$$[resp. \quad I = -\underbrace{(\theta - \theta_k^{est})^T P [\psi(x_k, \theta) - \psi(x_k, \theta_k^{est})]}_{> 0} < 0, \quad \forall \theta \neq \theta_k^{est}] \quad (3.19b)$$

On the other hand, the inequality $\|I\| \geq \|II\|$ can be simplified to the following inequality,

$$\|\theta_k^{est} - \theta\| \geq \|\beta(x_k) A(x_k)[\phi(x_k, \theta) - \phi(x_k, \theta_k^{est})]\| \quad (3.20)$$

and the above inequality can be re-written as:

$$\|(\theta_k^{est} - \theta)\| \geq \|\psi(x_k, \theta) - \psi(x_k, \theta_k^{est})\| \quad (3.21)$$

Since $\beta(x_k)$ is selected such that Assumption 2 is satisfied (Namely, $\psi(x_k, \theta)$ is Lipschitz with $L \leq 1$), the inequality (3.21) is satisfied. Hence, $z_{k+1}^T P(z_{k+1} - z_k) \leq 0$.

Secondly, let us prove (3.16b), namely, $z_k^T P(z_{k+1} - z_k) \leq 0$. By replacing each term in (3.16b) with its equivalent,

$$z_k^T P(z_{k+1} - z_k) = [\theta_k^{est} - \theta]^T P \left[\beta(x_k) A(x_k)[\phi(x_k, \theta) - \phi(x_k, \theta_k^{est})]\right] = I \quad (3.22)$$

is obtained. The right-hand side of (3.22) is equivalent to the term "$I$" in (3.22). By Assumption 1, the inequality (3.22) is satisfied. Hence, $z_k^T P(z_{k+1} - z_k) \leq 0$.

Thus, by showing that $\Delta V_{z_k} \leq 0$ (resp. $\Delta V_{z_k} < 0$), we proved that the estimation error dynamics (3.11) is Lyapunov stable (Lyapunov asymptotically stable if Assumption 1 and Assumption 2 are satisfied strictly).

**Remark 1**. Indeed, it is difficult to give a general exact construction for $\beta(x_k)$ that satisfies the stability conditions declared in the Proposition. The construction should be established for each system specifically. Nevertheless, we propose the following candidate structure for $\beta(x_k)$ to facilitate the construction where $\alpha \geq 1$ and $K(x_k, \theta_k^{est}) \in \Re^{s \times r}$ is selected such that Assumption 1 and Assumption 2 are satisfied.

$$\beta(x_k) = K(x_k, \theta_k^{est}) \frac{A(x_k)^T}{\alpha[\bar{\sigma}(A(x_k)^T A(x_k)) + 1]} \quad (3.23)$$

This structure is considered in the presented examples of this paper and it is shown that the structure is appropriate such that $K(x_k, \theta_k^{est})$ can be constructed where Assumption 1 and Assumption 2 are satisfied.

### 3.3 Discrete-time Adaptive IDA-PBC Control Design

In this subsection, firstly, the design of an adaptive IDA-PBC controller is presented, and then the stability analysis of the entire closed-loop system under the proposed I&I-based adaptive IDA-PBC is given. Let us consider the proposed uncertain system formulation (3.3), and recall the desired discrete-time Hamiltonian system given in (2.7). Equating the right-hand side of (3.3) to the right-hand side of (2.7) as follows:

$$\begin{aligned}[J(x_k) - R(x_k)] \Pi(x_k) \phi(x_k, \theta) + [J(x_k) - R(x_k)] \bar{\nabla} H_{kn}(x_k) + g(x_k) u_k \\ = J_d(x_k) \bar{\nabla} H_d(x_k)\end{aligned} \quad (3.24)$$

yields the term in the adaptive discrete-time control rule that re-shapes the system's energy as given below:

$$\begin{aligned}u_{es} = [g(x_k)^T g(x_k)]^{-1} g^T(x_k) \{J_d(x_k) \bar{\nabla} H_d(x_k) \\ - [J(x_k) - R(x_k)] \Pi(x_k) \phi(x_k, \theta_k^{est}) - [J(x_k) - R(x_k)] \bar{\nabla} H_{kn}(x_k)\}\end{aligned} \quad (3.25)$$

where $\phi(x_k, \theta_k^{est})$ is the function that contains the system's states together with the estimated values of the unknown parameters of the vector $\theta$. Since $H$ is shaped to the $H_d$ through $u_{es}$, the damping injection control rule is formed as follows:

$$u_{di} = -K_v \, g^T(x_k) \, \bar{\nabla} H_d(x_k) \tag{3.26}$$

Therefore, the total discrete-time IDA-PBC control rule is obtained as the sum of (3.25) and (3.26) where this control rule has been derived under the assumption that the desired continuous-time closed-loop system is available which is difficult to be constructed for the underactuated non-linear Hamiltonian systems such that the matching condition in (2.8) is satisfied.

Now, let us move to the stability analysis of the closed-loop system, we claim in the following theorem that the proposed adaptive IDA-PBC controller with the proposed I&I estimator ensures the closed-loop system is locally asymptotically stable in a sufficiently large set around the origin and the adaptive controller eventually assigns the desired energy function to the system.

**Theorem.** Consider the uncertain discrete-time non-linearly parameterized Hamiltonian system given in (3.3) where the unknown system parameters vector "$\theta$" is constant or varies much slower than the estimator dynamics. Let the parameter estimation error be defined as (3.7), and the free design function "$\beta(x_k)$" is selected as given in (3.23) such that $\psi(x_k, \theta)$ in (3.12) is a strictly P-monotone and Lipschitz with $L < 1$ (Assumption 1 and Assumption 2 are strictly satisfied), and the parameters update law is given as in (3.10). Then, the adaptive control rule stated in (3.25) to (3.26) eventually assigns the desired energy function to the system (3.3) and makes the closed-loop system locally asymptotically stable in any sufficiently large set.

**Proof.** Let's recall the system dynamic equations with energy function uncertainties,

$$x_{k+1} - x_k = T(J - R)\,\Pi(x_k)\phi(x_k, \theta) + T(J - R)\bar{\nabla} H_{kn} + T\, g\, u_k \tag{3.27}$$

and recall that the control signal is the sum of (3.25) and (3.26) as follows,

$$u_k = (g^T g)^{-1} g^T \{J_d \, \bar{\nabla} H_d - [J - R]\,\Pi(x_k)\phi(x_k, \theta_k^{est}) - [J - R]\bar{\nabla} H_{kn}\} - K_v g^T \, \bar{\nabla} H_d \tag{3.28}$$

By replacing $u_k$ in (3.27) by its equivalent in (3.28), we obtain

$$\begin{aligned} x_{k+1} - x_k &= \underbrace{T[\mathcal{G}J_d - R_d]}_{B} \bar{\nabla} H_d + \underbrace{T[I_n - \mathcal{G}][J - R]}_{C} \bar{\nabla} H_{kn} + \underbrace{T[J - R]\Pi(x_k)}_{A(x)} \phi(x_k, \theta) \\ &\quad - \mathcal{G}\underbrace{T[J - R]\Pi(x_k)}_{A(x)} \phi(x_k, \theta_k^{est}) \\ &= B\bar{\nabla} H_d + C\bar{\nabla} H_{kn} + A(x_k)\phi(x_k, \theta) - \mathcal{G}A(x_k)\phi(x_k, \theta_k^{est}). \end{aligned} \tag{3.29}$$

with

$$\mathcal{G} = g\,(g^T g)^{-1} g^T$$

For the closed-loop system, let's select the following function as the candidate Lyapunov function

$$V_k = \sigma\, \tilde{H}_d(x_k) + \gamma\, V_{z_k} \tag{3.30}$$

where $\tilde{H}_d(x_k) = H_d(x_k) + \in(x_k)$, $\in(x_k)$ denotes the uncertainty-induced energy function variation from the desired energy function, and where σ and γ are free parameters. Then, the time difference of $V_k$ is

$$\Delta V_k = V_{k+1} - V_k = \sigma\left[\widetilde{H}_d(x_{k+1}) - \widetilde{H}_d(x_k)\right] + \gamma\left[z_{k+1}^T P z_{k+1} - z_k^T P z_k\right]. \tag{3.31}$$

Here, we can write the following relations considering discrete-gradient conditions given in (2.2a),

$$\widetilde{H}_d(x_{k+1}) - \widetilde{H}_d(x_k) = \bar{\nabla}\widetilde{H}_d^T (x_{k+1} - x_k) \tag{3.32}$$

By multiplying both sides of (3.29) by $\bar{\nabla}\widetilde{H}_d^T$, we obtain

$$\bar{\nabla}\widetilde{H}_d^T (x_{k+1} - x_k) = \bar{\nabla}\widetilde{H}_d^T \left(B\bar{\nabla}H_d + C\bar{\nabla}H_{kn} + A(x_k)\phi(x_k,\theta) - \mathcal{G}A(x_k)\phi(x_k,\theta_k^{est})\right) \tag{3.33}$$

and by equating (3.33) with (3.32), we obtain

$$\begin{aligned}\widetilde{H}_d(x_{k+1}) - \widetilde{H}_d(x_k) \\ = \bar{\nabla}\widetilde{H}_d^T B\bar{\nabla}H_d + \bar{\nabla}\widetilde{H}_d^T \left(C\bar{\nabla}H_{kn} + A(x_k)\phi(x_k,\theta) - \mathcal{G}A(x_k)\phi(x_k,\theta_k^{est})\right)\end{aligned} \tag{3.34}$$

Furthermore, by replacing $\bar{\nabla}\widetilde{H}_d^T$ and B in the term $\bar{\nabla}\widetilde{H}_d^T B\bar{\nabla}H_d$ of the above equation with their equivalents,

$$\begin{aligned}\widetilde{H}_d(x_{k+1}) - \widetilde{H}_d(x_k) \\ = \bar{\nabla}H_d^T T\mathcal{G}J_d\bar{\nabla}H_d - \bar{\nabla}H_d^T T R_d\bar{\nabla}H_d + \bar{\nabla}\in^T (x_k)T[\mathcal{G}J_d - R_d]\bar{\nabla}H_d \\ + \bar{\nabla}\widetilde{H}_d^T\left(C\bar{\nabla}H_{kn} + A(x_k)\phi(x_k,\theta) - \mathcal{G}A(x_k)\phi(x_k,\theta_k^{est})\right)\end{aligned} \tag{3.35}$$

can be obtained. Then, by utilizing (3.18), (3.22), and (3.35) in (3.31),

$$\begin{aligned}\Delta V_k = \sigma\bar{\nabla}H_d^T T\mathcal{G}J_d\bar{\nabla}H_d - \sigma\bar{\nabla}H_d^T T R_d\bar{\nabla}H_d + \sigma\bar{\nabla}\in^T (x_k)T[\mathcal{G}J_d - R_d]\bar{\nabla}H_d \\ + \sigma\bar{\nabla}\widetilde{H}_d^T\left(C\bar{\nabla}H_{kn} + A(x_k)\phi(x_k,\theta) - \mathcal{G}A(x_k)\phi(x_k,\theta_k^{est})\right) \\ + \gamma\,[2I + II]\end{aligned} \tag{3.36}$$

and

$$\begin{aligned}\Delta V_k = -\Bigg(\sigma\underbrace{\bar{\nabla}H_d^T T R_d\bar{\nabla}H_d}_{F>0} - \gamma\underbrace{[2I + II]}_{Q<0}\Bigg) + \underbrace{\sigma\bar{\nabla}H_d^T T}_{a_1}\underbrace{\mathcal{G}J_d\bar{\nabla}H_d}_{a_2} \\ + \underbrace{\sigma\bar{\nabla}\in^T (x_k)T}_{b_1}\underbrace{[\mathcal{G}J_d - R_d]\bar{\nabla}H_d}_{b_2} \\ + \underbrace{\sigma\bar{\nabla}\widetilde{H}_d^T}_{c_1}\underbrace{[C\bar{\nabla}H_{kn} + A(x_k)\phi(x_k,\theta) - \mathcal{G}A(x_k)\phi(x_k,\theta_k^{est})]}_{c_2}\end{aligned} \tag{3.37}$$

can be written. Using Young's inequality ($ab \leq \frac{a^2}{2\varepsilon} + \frac{\varepsilon b^2}{2}$), the terms indicated as $a_i$, $b_i$ and $c_i$ $(i = 1, 2)$ in (3.37) can be substituted with larger positive definite terms as follows:

$$\Delta V_k \leq \underbrace{-(\sigma F - \gamma Q)}_{\Lambda_1 < 0} + \underbrace{\frac{a_1^T a_1}{2\varepsilon_a} + \frac{\varepsilon_a a_2^T a_2}{2} + \frac{b_1^T b_1}{2\varepsilon_b} + \frac{\varepsilon_b b_2^T b_2}{2} + \frac{c_1^T c_1}{2\varepsilon_c} + \frac{\varepsilon_c c_2^T c_2}{2}}_{\Lambda_2} \tag{3.38}$$

At this stage, it can be stated that some $\sigma$ and $\gamma$ always exist in such a way that,

$$\Lambda_1 + \Lambda_2 < 0 \tag{3.39}$$

namely,

$$\Delta V_k < 0 \tag{3.40}$$

is satisfied in an arbitrarily large local set around the origin. Thus all the closed-loop system trajectories are bounded and locally asymptotically stable in the sense of Lyapunov.

**Remark 2**. In the above theorem, we proved that the adaptive controller renders the closed-loop system locally asymptotically stable. That is to say in the transient of the estimator, the controller may not satisfy the matching condition in (2.8) but it keeps the system stable. Moreover, in case of Assumptions 1 and 2 are strictly satisfied as our theorem is assumed; the proposed estimator is asymptotically stable and thus it will converge to the true values of the unknown parameters. Remember that, we assumed that the desired system satisfies the matching condition with true system parameters. Namely, it is designed in terms of system parameters such that it satisfies the matching condition. Therefore, our adaptive controller will satisfy the matching condition and assign the desired energy function after the estimator converges to the true parameters. On the other hand, the mentioned matching condition is satisfied generally for the fully actuated systems.

The entire structure of the proposed adaptive controller is shown in Fig. 1.

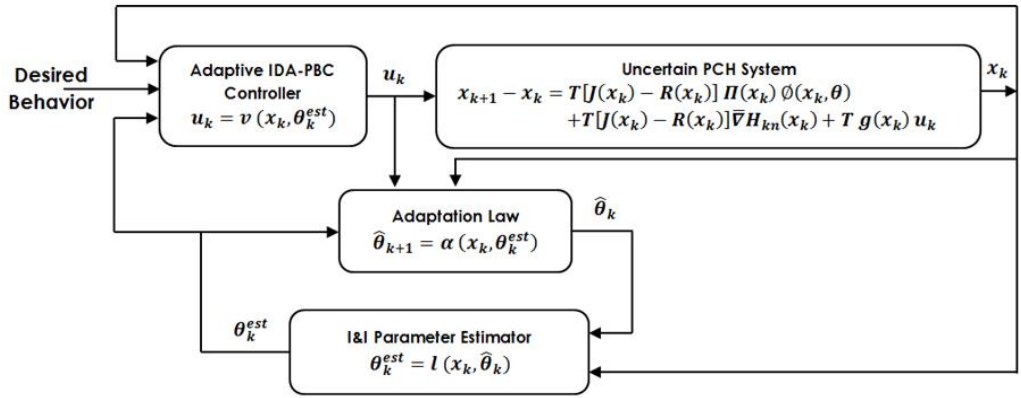

**FIGURE 1** The generic structure of the proposed Adaptive IDA-PBC Control

## 4. EXAMPLES AND SIMULATIONS

In this section, the proposed adaptive IDA-PBC method is applied to two nonlinearly parameterized Hamiltonian systems, and the performance of the method is tested by simulations. The desired Hamiltonian systems for the examples are taken from the continuous-time IDA-PBC controller constructions in the literature. Namely, the considered desired Hamiltonian systems satisfy the matching condition given in (2.8) in the case of using the true parameters in the controller.

### 4.1 The Single Pendulum System

Consider the energy function of the Hamiltonian system as:

$$H(q,p) = K(q,p) + V(q) = \frac{1}{2}p^T M^{-1}(q)p + V(q) = \frac{1}{2}\underbrace{\left(mL^2\right)}_{M}^{-1} p^2 \underbrace{-mgL\cos q}_{V(q)} \quad (4.1)$$

where the system is given in the standard coordinates, $x = [q \quad p]^T$ in (2.1), $V(q)$ is the potential energy, $K(q,p)$ is the kinetic energy, $q$ is the generalized position, $p$ is the generalized momentum, and $M(q)$ is the generalized inertia. The planar pendulum is a particle of mass $m$ in a constant gravitational field $g$, that is attached to a rigid, massless rod of length $L$, and $\theta$ is the vertical angle of the pendulum as shown in Fig. 2.

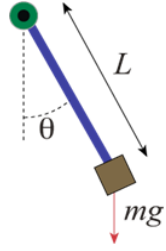

**FIGURE 2** The single pendulum system

The gradient of the energy function given in (4.1) is given below.

$$\nabla H = \begin{bmatrix}\nabla_q H \\ \nabla_p H\end{bmatrix} = \begin{bmatrix} mgL\sin q \\ \frac{1}{mL^2}p \end{bmatrix} \quad (4.2)$$

The port-controlled Hamiltonian model of the system is as given below where no dissipative effects are considered.

$$\begin{bmatrix}\dot{q}\\\dot{p}\end{bmatrix} = (J-R)\begin{bmatrix}\nabla_q H\\\nabla_p H\end{bmatrix} + \begin{bmatrix}0\\G\end{bmatrix}u(k) \quad (4.3a)$$

$$J = \begin{bmatrix} 0 & 1 \\ -1 & 0 \end{bmatrix}, \quad R = 0 \in \Re^{2\times 2}, \quad G = 1 \quad (4.3b)$$

Here, the desired Hamiltonian energy function that has been designed in [30] and restated in [61] is used:

$$H_d(q,p) = K_d(q,p) + V_d(q) = \frac{1}{2}p^T M_d^{-1}(q)p + V_d(q)$$

$$H_d(q,p) = \frac{1}{2}\underbrace{\left(mL^2\right)}_{M_d}^{-1} p^2 \underbrace{-mgL\cos q + \frac{1}{2}k_p\left(q - q^* - \frac{1}{k_p}mgL\sin q^*\right)^2}_{V_d(q)} \quad (4.4)$$

where $q^*$ is the desired position and $k_p > mgL$ is chosen to satisfy the minimum of $H_d$ at $q^*$ [61]. Therefore the gradient of the desired energy function given in (4.4) is:

$$\nabla H_d = \begin{bmatrix} \nabla_q H_d \\ \nabla_p H_d \end{bmatrix} = \begin{bmatrix} mgL \sin q + k_p \left( q - q^* - \frac{1}{k_p} mgL \sin q^* \right) \\ \frac{1}{mL^2} p \end{bmatrix} \quad (4.5)$$

According to [42] and [28], the desired continuous Hamiltonian system is considered as:

$$\begin{bmatrix} \dot{q} \\ \dot{p} \end{bmatrix} = (J_d - R_d) \begin{bmatrix} \nabla_q H_d \\ \nabla_p H_d \end{bmatrix} \quad (4.6a)$$

$$J_d = \begin{bmatrix} 0 & M^{-1} M_d \\ -M_d M^{-1} & J_2 \end{bmatrix}, \quad R_d = \begin{bmatrix} 0 & 0 \\ 0 & G K_v G^T \end{bmatrix} \quad (4.6b)$$

where $J_2(q,p) = -J_2^T(q,p)$, $R_d(q,p) = R_d^T(q,p)$, $K_v = K_v^T \geq 0$.

Note that, if $M(q) = M \in \Re^{n \times n}$, the system is called a "Separable Hamiltonian System"; otherwise, it is called a "Non-separable Hamiltonian System". Since the system is separable, namely matrix M in (4.1) is independent of $q$, $J_2 = 0$ is considered in the matrix $J_d$ [62].

We can obtain the discrete-time gradient of energy function (4.1) considering (2.6) as follows:

$$\overline{\nabla} H = \begin{bmatrix} \overline{\nabla}_{\bar{q}} H \\ \overline{\nabla}_{\bar{p}} H \end{bmatrix} = \begin{bmatrix} mgL \sin \bar{q} \\ \frac{1}{mL^2} \bar{p} \end{bmatrix} \quad (4.7)$$

where, $\bar{q} = \frac{3q_k - q_{k-1}}{2}$, $\bar{p} = \frac{3p_k - p_{k-1}}{2}$ and $q_k, p_k$ are the values of the system states at $k^{th}$ sampling instance, (i.e., Eq. (4), $x_k = [q_k \quad p_k]^T$). Assume that the uncertainties are considered in the system's parameter $L$. As a result; the discrete gradient of the system's energy function could be parameterized non-linearly as follows:

$$\overline{\nabla} H = \underbrace{\begin{bmatrix} 0 \\ 0 \end{bmatrix}}_{\overline{\nabla} H_{kn}} + \underbrace{\begin{bmatrix} mgL \sin \bar{q} \\ \frac{1}{mL^2} \bar{p} \end{bmatrix}}_{\overline{\nabla} \tilde{H}_{un}} = \underbrace{\begin{bmatrix} mgL \sin \bar{q} \\ \frac{1}{mL^2} \bar{p} \end{bmatrix}}_{\overline{\nabla} \tilde{H}_{un}} = \underbrace{\begin{bmatrix} mg \sin \bar{q} & 0 \\ 0 & \frac{1}{m} \bar{p} \end{bmatrix}}_{\Pi(x_k)} \underbrace{\begin{bmatrix} L \\ 1/L^2 \end{bmatrix}}_{\phi(x_k, \theta)} \quad (4.8)$$

Namely, the discrete gradient of the system's energy function can be brought into the following form;

$$\overline{\nabla} \tilde{H}_{un} = \Pi(x_k) \phi(x_k, \theta) \quad (4.9)$$

where the unknown parameter is,

$$\theta = L \quad (4.10)$$

### 4.1.1 Construction of the free design function of the estimator

By Proposition, the free design function $\beta(x_k)$ to be selected such that both the conditions are satisfied as assumed in Assumption 1 and Assumption 2 respectively, namely such that $\psi(x_k, \theta)$ defined in (3.12) is a P-monotone and Lipschitz function with $L < 1$.

Firstly, let us consider the first condition:

$$(\theta - \theta_k^{est})^T \beta(x_k) A(x_k) [\phi(x_k, \theta) - \phi(x_k, \theta_k^{est})] \geq 0 \quad (4.11)$$

Since $s = 1$ and $r = 2$, then $K(x_k, \theta_k^{est}) \in \Re^{1 \times 2}$ in (3.23) as restated below,

$$\beta(x_k) = K(x_k, \theta_k^{est}) \frac{A(x_k)^T}{\alpha[\bar{\sigma}(A(x_k)^T A(x_k)) + 1]} = [K_1 \quad K_2] \frac{A(x_k)^T}{\alpha[\bar{\sigma}(A(x_k)^T A(x_k)) + 1]} \quad (4.12)$$

where $K_i(x_k, \theta_k^{est})$ to be selected such that $\beta(x_k)$ satisfies (4.11) as shown in the following steps. Let us construct $A(x_k)$ as below:

$$A(x_k) = T[J - R]\Pi(x_k) = \begin{bmatrix} 0 & \frac{T\bar{p}}{m} \\ -Tmg \sin \bar{q} & 0 \end{bmatrix} \quad (4.13)$$

In accordance,

$$\phi(x_k, \theta) - \phi(x_k, \theta_k^{est}) = \begin{bmatrix} \theta \\ 1/\theta^2 \end{bmatrix} - \begin{bmatrix} \theta_{est} \\ 1/\theta_{est}^2 \end{bmatrix} = \begin{bmatrix} 1 \\ \frac{-(\theta + \theta_{est})}{\theta^2 \theta_{est}^2} \end{bmatrix} [\theta - \theta_{est}] \quad (4.14)$$

is obtained. Replacing all the terms of (4.11) by their equivalents yields:

$$\equiv [\theta - \theta_{est}]^T [K_1 \quad K_2] A(x_k)^T A(x_k) \begin{bmatrix} 1 \\ \frac{-(\theta + \theta_{est})}{\theta^2 \theta_{est}^2} \end{bmatrix} [\theta - \theta_{est}]$$

$$\equiv [\theta - \theta_{est}]^T [K_1 \quad K_2] \begin{bmatrix} T^2 m^2 g^2 \sin^2 \bar{q} & 0 \\ 0 & \frac{T^2 \bar{p}^2}{m^2} \end{bmatrix} \begin{bmatrix} 1 \\ \frac{-(\theta + \theta_{est})}{\theta^2 \theta_{est}^2} \end{bmatrix} [\theta - \theta_{est}] \quad (4.15)$$

$$\equiv \left( K_1 [T^2 m^2 g^2 \sin^2 \bar{q}] - K_2 \left[ \frac{(\theta + \theta_k^{est})}{\theta^2 \theta_k^{est \, 2}} \left( \frac{T^2 \bar{p}^2}{m^2} \right) \right] \right) [\theta - \theta_{est}]^2$$

Now, let us select $K_1 = c_1$ and $K_2 = 0$, namely, $\beta(x_k)$ as follows:

$$\beta(x_k) = [c_1 \quad 0] \frac{A(x_k)^T}{\alpha[\bar{\sigma}(A(x_k)^T A(x_k)) + 1]} \quad (4.16)$$

where $c_1 > 0$ is an arbitrary free positive scalar. Then, (4.15) becomes:

$$\frac{c_1 (T^2 m^2 g^2 \sin^2 \bar{q})}{\alpha[\bar{\sigma}(A(x_k)^T A(x_k)) + 1]} [\theta - \theta_{est}]^2 \geq 0 \quad (4.17)$$

Hence, the first condition is satisfied.
Secondly, let us now consider the second condition:

$$\|(\theta_k^{est} - \theta)\| \geq \|\beta(x_k) A(x_k) [\phi(x_k, \theta) - \phi(x_k, \theta_k^{est})]\| \quad (4.18)$$

Again, replacing all the terms of (4.18) with their equivalents yields:

$$\|(\theta_{est} - \theta)\| \geq \left\| \frac{c_1(T^2 m^2 g^2 sin^2 \bar{q})}{\alpha[\bar{\sigma}(A(x_k)^T A(x_k)) + 1]} (\theta - \theta_{est}) \right\| \tag{4.19}$$

Noting that if the following inequality is satisfied, then inequality (4.19) is satisfied as well, which means that $\psi(x_k, \theta)$ is locally Lipschitz with $L < 1$.

$$\left\| \frac{c_1(T^2 m^2 g^2 sin^2 \bar{q})}{\alpha[\bar{\sigma}(A(x_k)^T A(x_k)) + 1]} \right\| < 1 \tag{4.20}$$

Thus, the above inequality is reduced to the following relation

$$\frac{\alpha}{c_1} > \frac{(T^2 m^2 g^2 sin^2 \bar{q})}{[\bar{\sigma}(A(x_k)^T A(x_k)) + 1]} \tag{4.21}$$

Therefore, to satisfy the condition strictly, the estimator parameter can be selected as follows where $\delta > 0$ is a positive free constant.

$$c_1 = \frac{[\bar{\sigma}(A(x_k)^T A(x_k)) + 1]}{(T^2 m^2 g^2 sin^2 \bar{q}) + \delta} \alpha$$

Note that, now the proposed adaptive IDA-PBC controller in (3.28) can be straightforwardly constructed by making use of the derived non-linearly parameterized discrete-time gradient of the uncertain energy function of the system, the desired energy function and it's discrete-gradient, and the selected design function $\beta(x_k)$ above.

### 4.1.2 Simulation results

In the simulations, the desired energy function parameters are taken as: $k_d = 40$, the design parameter of the damping injection control rule is chosen as $K_v = 5$, and the parameters of $\beta(x)$ are selected as $k_1 = 100$ and $\alpha = 2$. The simulations are carried out with sampling time $T_s = 0.01s$. The initial conditions are selected as $[q(0) \quad p(0)]^T = [0.7 \quad 0.5]^T$ and $[\theta_{est}(0)] = [0.01]$. The nominal value of the system parameter is taken as $m = 1kg$ and $L = 4m$. The desired angle position is taken as $q^* = 2 \, rad$.

The simulation results for this example are given in Figs. 3-5. The closed-loop system dynamics $(q, p)$ for three different cases are illustrated together with desired system dynamics in Fig. 3. In the figure, the dynamics of the desired system are depicted in black color. The red color depicts the closed-loop system dynamics for the case where the actual system parameter is different from the nominal value of the system parameter, namely $L = 2$, and the system is controlled by a non-adaptive controller such that the actual system parameter value is not known by the non-adaptive controller, and thus the non-adaptive controller uses the nominal value of system parameter instead, namely, $L = 4$. The blue color corresponds to the case where the actual system parameter is $L = 2$ and the system is controlled by the proposed adaptive controller that uses the estimated parameter obtained by the proposed parameter estimator. Lastly, the green color depicts the closed-loop system dynamics for the case where the system is controlled by the controller that knows and uses the actual system parameter, namely, $L = 2$. The figure illustrates that the adaptive controller preserves the closed-loop performance under uncertainty successfully while the non-adaptive controller fails to maintain the closed-loop performance.

In Fig. 4, the dynamic of the estimated parameter is given. It can be seen in the figure that the estimated parameter successfully converges to the true parameter. In Fig. 5, the corresponding control signals are shown for the presented three cases.

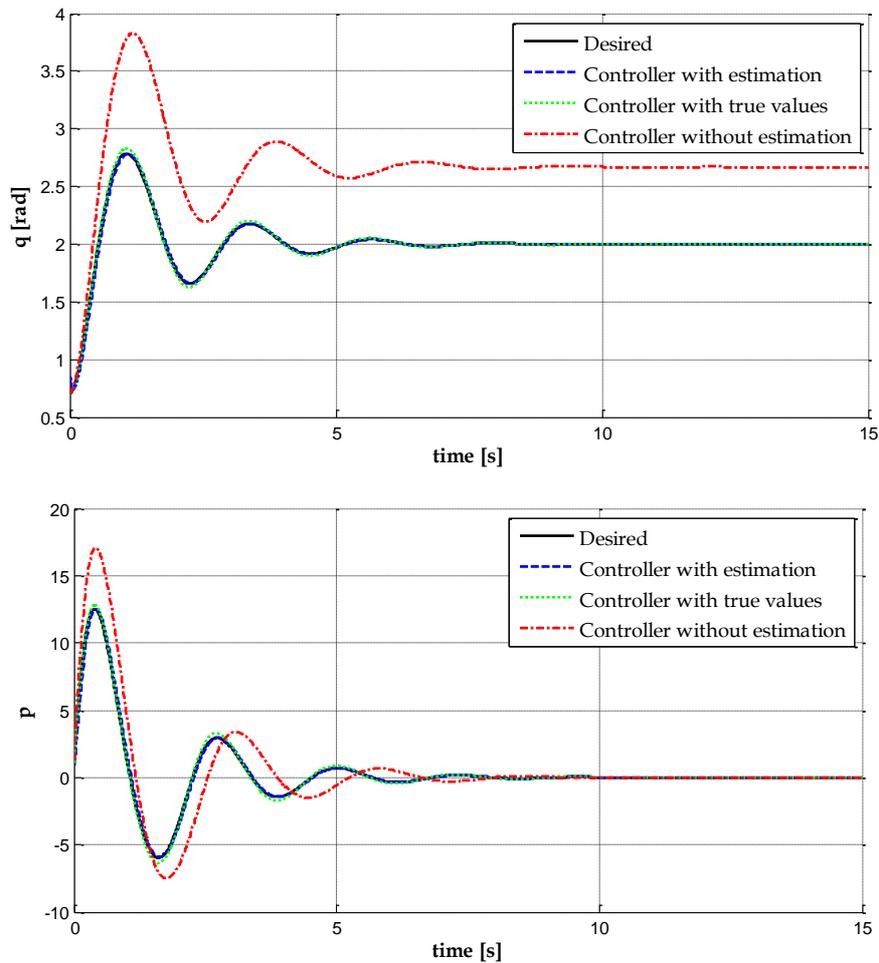

**FIGURE 3**  The closed-loop dynamics under control of adaptive and non-adaptive controllers

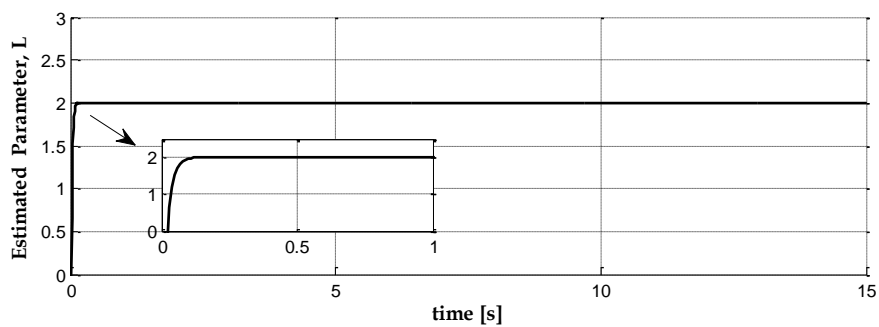

**FIGURE 4**  The dynamic of the estimated parameter

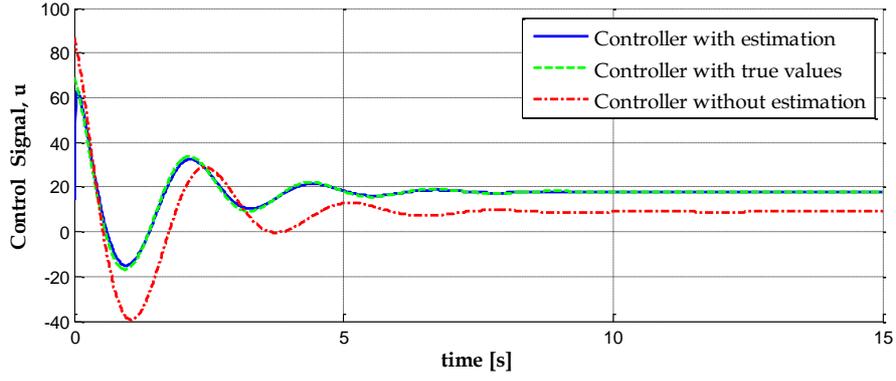

**FIGURE 5** The control signals for the different considered cases

4.2 The Single Inertia Wheel Pendulum System

Here, the inertia wheel pendulum system shown in Fig. 6 consists of a pendulum with a balanced rotor at the end is considered. The motor torque produces an angular acceleration in the wheel which generates a coupling torque at the pendulum axis.

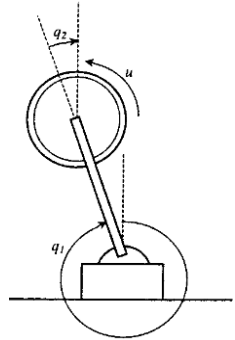

**FIGURE 6** The inertia wheel pendulum system [28]

The Hamiltonian energy function of the system is [28]:

$$H(q,p) = K(q,p) + V(q)$$

$$H(q,p) = \frac{1}{2}p^T \underbrace{\begin{bmatrix} I_1 & 0 \\ 0 & I_2 \end{bmatrix}^{-1}}_{M} p + \underbrace{m_3(\cos q_1 - 1)}_{V(q)} = \frac{1}{2I_1}p_1^2 + \frac{1}{2I_2}p_2^2 + m_3(\cos q_1 - 1) \quad (4.22)$$

where the system is given in the standard coordinates, $x = \begin{bmatrix} q & p \end{bmatrix}^T$ in (2.1), $V(q)$ is the potential energy, $K(q,p)$ is the kinetic energy, $q$ is the vector of the generalized positions, $p$ is the vector of the generalized momentums, $M(q)$ is the symmetric and the positive generalized inertia matrix, $I_1$ and $I_2$ are positive constant parameters, $m_3 = mgL$ is a positive parameter such that $m$ is the pendulum mass, $L$ is the pendulum length, and $g$ is the gravity constant. Besides, the equilibrium position to be stabilized are the upward position $q_1^* = 0$ and the inertia disk aligned $q_2^* = 0$.

The gradient of the energy function (4.22) is as follows:

$$\nabla H = \begin{bmatrix} \nabla_{q_1} H \\ \nabla_{q_2} H \\ \nabla_{p_1} H \\ \nabla_{p_2} H \end{bmatrix} = \begin{bmatrix} -m_3 \sin q_1 \\ 0 \\ \frac{1}{I_1} p_1 \\ \frac{1}{I_2} p_2 \end{bmatrix} \quad (4.23)$$

The port-controlled Hamiltonian model of the system is as given below where no dissipative effects are considered.

$$\begin{bmatrix} \dot{q} \\ \dot{p} \end{bmatrix} = (J - R) \begin{bmatrix} \nabla_q H \\ \nabla_p H \end{bmatrix} + \begin{bmatrix} 0 \\ G \end{bmatrix} u(k) \quad (4.24a)$$

$$J = \begin{bmatrix} 0 & I_n \\ -I_n & 0 \end{bmatrix} \in \Re^{4 \times 4}, \quad R = 0 \in \Re^{4 \times 4}, \quad G = [-1 \quad 1]^T \quad (4.24b)$$

Here, the desired Hamiltonian energy function that has been designed in [30], and restated in [61] and [63] is used:

$$H_d(q,p) = K_d(q,p) + V_d(q) = \frac{1}{2} p^T M_d^{-1}(q) p + V_d(q)$$

$$H_d(q,p) = \frac{1}{2} [p_1 \quad p_2] \underbrace{\begin{bmatrix} a_1 & a_2 \\ a_2 & a_3 \end{bmatrix}^{-1}}_{M_d} \begin{bmatrix} p_1 \\ p_2 \end{bmatrix} + \underbrace{\frac{I_1 m_3}{a_1 + a_2} (\cos q_1 - 1) + \frac{k_1}{2} (q_2 + \gamma_2 q_1)^2}_{V_d(q)} \quad (4.25)$$

$$H_d(q,p) = \frac{1}{2k_2} (a_3 p_1^2 - 2 a_2 p_1 p_2 + a_1 p_2^2) + \frac{I_1 m_3}{a_1 + a_2} (\cos q_1 - 1) + \frac{k_1}{2} (q_2 + \gamma_2 q_1)^2$$

where $k_1 > 0$ is an adjustable parameter, $\gamma_2 = -I_1(a_2 + a_3)/I_2(a_1 + a_2)$, and $k_2 = a_1 a_3 - a_2^2$. In addition, $a_1$, $a_2$, and $a_3$ should satisfy the following inequalities:

$$a_1 > 0, \quad a_1 a_3 > a_2^2, \quad a_1 + a_2 < 0 \quad (4.26)$$

Therefore, the gradient of the desired energy function is:

$$\nabla H_d = \begin{bmatrix} \nabla_{q_1} H_d \\ \nabla_{q_2} H_d \\ \nabla_{p_1} H_d \\ \nabla_{p_2} H_d \end{bmatrix} = \begin{bmatrix} -\frac{I_1 m_3}{a_1 + a_2} \sin q_1 + k_1 \gamma_2 (q_2 + \gamma_2 q_1) \\ k_1 (q_2 + \gamma_2 q_1) \\ (a_3 p_1 - a_2 p_2)/k_2 \\ (a_1 p_2 - a_2 p_1)/k_2 \end{bmatrix} \quad (4.27)$$

According to [42] and [28], the desired Hamiltonian system is considered as:

$$\begin{bmatrix} \dot{q} \\ \dot{p} \end{bmatrix} = (J_d - R_d) \begin{bmatrix} \nabla_q H_d \\ \nabla_p H_d \end{bmatrix} \quad (4.28a)$$

$$J_d = \begin{bmatrix} 0 & M^{-1}M_d \\ -M_d M^{-1} & J_2 \end{bmatrix}, \quad R_d = \begin{bmatrix} 0 & 0 \\ 0 & GK_v G^T \end{bmatrix} \qquad (2.28b)$$

where $J_2(q,p) = -J_2^T(q,p)$, $R_d(q,p) = R_d^T(q,p)$, $K_v = K_v^T \geq 0$. Since the system is separable, namely matrix $M$ in (4.22) is independent of $q$, then $J_2 = 0$ is considered in the matrix $J_d$ [62].

Referring to (2.6), the discrete gradient of (4.22) is obtained and separated into its known and unknown parts as follows by assuming that the uncertainties are considered in the system's parameters $I_1$ and $I_2$.

$$\bar{\nabla} H = \begin{bmatrix} \bar{\nabla}_{\bar{q}_1} H \\ \bar{\nabla}_{\bar{q}_2} H \\ \bar{\nabla}_{\bar{p}_1} H \\ \bar{\nabla}_{\bar{p}_2} H \end{bmatrix} = \begin{bmatrix} -m_3 \sin \bar{q}_1 \\ 0 \\ \dfrac{1}{I_1}\bar{p}_1 \\ \dfrac{1}{I_2}\bar{p}_2 \end{bmatrix} = \underbrace{\begin{bmatrix} -m_3 \sin \bar{q}_1 \\ 0 \\ 0 \\ 0 \end{bmatrix}}_{\bar{\nabla} H_{kn}} + \underbrace{\begin{bmatrix} 0 \\ 0 \\ \dfrac{1}{I_1}\bar{p}_1 \\ \dfrac{1}{I_2}\bar{p}_2 \end{bmatrix}}_{\bar{\nabla}\tilde{H}_{un}} \qquad (4.29)$$

where, $\bar{q}_i = \dfrac{3q_{i_k} - q_{i_{k-1}}}{2}$, $\bar{p}_i = \dfrac{3p_{i_k} - p_{i_{k-1}}}{2}$, $(i = 1,2)$ and $q_{i_k}, p_{i_k}$ are the values of the system states at $k^{th}$ sampling instance, (i.e., Eq. (4), $x_k = [q_{1_k} \ q_{2_k} \ p_{1_k} \ p_{2_k}]^T$). Hence, the unknown part $\bar{\nabla}\tilde{H}_{un}$ can be non-linearly parameterized as follows,

$$\bar{\nabla}\tilde{H}_{un} = \begin{bmatrix} 0 \\ 0 \\ \dfrac{1}{I_1}\bar{p}_1 \\ \dfrac{1}{I_2}\bar{p}_2 \end{bmatrix} = \underbrace{\begin{bmatrix} 0 & 0 \\ 0 & 0 \\ \bar{p}_1 & 0 \\ 0 & \bar{p}_2 \end{bmatrix}}_{\Pi(x_k)} \underbrace{\begin{bmatrix} \dfrac{1}{I_1} \\ \dfrac{1}{I_2} \end{bmatrix}}_{\phi(x_k,\theta)} \qquad (4.30)$$

and compactly formulated as below:

$$\bar{\nabla}\tilde{H}_{un} = \Pi(x_k)\,\phi(x_k, \theta) \qquad (4.31)$$

where

$$\theta = [\theta_1 \ \theta_2]^T = [I_1 \ I_2]^T. \qquad (4.32)$$

### 4.2.1 Construction of the free design function of the estimator

By Proposition, the free design function $\beta(x_k)$ to be selected such as both the conditions are satisfied as assumed in Assumption 1 and Assumption 2 respectively, namely such that $\psi(x_k, \theta)$ defined in (3.12) is a P-monotone and Lipschitz function with $L < 1$.

Firstly, let us consider the first condition:

$$(\theta - \theta_k^{est})^T \beta(x_k) A(x_k) [\phi(x_k, \theta) - \phi(x_k, \theta_k^{est})] \geq 0 \qquad (4.33)$$

Since $s = 2$ and $r = 2$, then $K(x_k, \theta_k^{est}) \in \Re^{2 \times 2}$ in (3.23) as restated below,

$$\beta(x_k) = K(x_k, \theta_k^{est}) \frac{A(x_k)^T}{\alpha[\bar{\sigma}(A(x_k)^T A(x_k)) + 1]} = \begin{bmatrix} K_1 & 0 \\ 0 & K_2 \end{bmatrix} \frac{A(x_k)^T}{\alpha[\bar{\sigma}(A(x_k)^T A(x_k)) + 1]} \quad (4.34)$$

where $K_i(x_k, \theta_k^{est})$ to be selected such that $\beta(x_k)$ satisfies (4.33) as shown in the following steps. Let us construct $A(x_k)$ as below:

$$A(x_k) = T[J - R]\Pi(x_k) = \begin{bmatrix} T\bar{p}_1 & 0 \\ 0 & T\bar{p}_2 \\ 0 & 0 \\ 0 & 0 \end{bmatrix} \quad (4.35)$$

In accordance,

$$\phi(x_k, \theta) - \phi(x_k, \theta_k^{est}) = \begin{bmatrix} \frac{1}{\theta_1} \\ \frac{1}{\theta_2} \end{bmatrix} - \begin{bmatrix} \frac{1}{\theta_1^{est}} \\ \frac{1}{\theta_2^{est}} \end{bmatrix} = \begin{bmatrix} \frac{-1}{\theta_1 \theta_1^{est}} & 0 \\ 0 & \frac{-1}{\theta_2 \theta_2^{est}} \end{bmatrix} \underbrace{\begin{bmatrix} \theta_1 - \theta_1^{est} \\ \theta_2 - \theta_2^{est} \end{bmatrix}}_{[\theta - \theta_k^{est}]} \quad (4.36)$$

is obtained. Replacing all the terms of (4.33) by their equivalent yields:

$$\equiv [\theta - \theta_k^{est}]^T \begin{bmatrix} K_1 & 0 \\ 0 & K_2 \end{bmatrix} A(x_k)^T A(x_k) \begin{bmatrix} \frac{-1}{\theta_1 \theta_1^{est}} & 0 \\ 0 & \frac{-1}{\theta_2 \theta_2^{est}} \end{bmatrix} [\theta - \theta_k^{est}]$$

$$\equiv [\theta - \theta_k^{est}]^T \begin{bmatrix} K_1 & 0 \\ 0 & K_2 \end{bmatrix} \begin{bmatrix} T^2\bar{p}_1^2 & 0 \\ 0 & T^2\bar{p}_2^2 \end{bmatrix} \begin{bmatrix} \frac{-1}{\theta_1 \theta_1^{est}} & 0 \\ 0 & \frac{-1}{\theta_2 \theta_2^{est}} \end{bmatrix} [\theta - \theta_k^{est}] \quad (4.37)$$

$$\equiv [\theta - \theta_k^{est}]^T \begin{bmatrix} -K_1 \frac{T^2\bar{p}_1^2}{\theta_1 \theta_1^{est}} & 0 \\ 0 & -K_2 \frac{T^2\bar{p}_2^2}{\theta_2 \theta_2^{est}} \end{bmatrix} [\theta - \theta_k^{est}]$$

$$\equiv -K_1 \frac{T^2\bar{p}_1^2}{\theta_1 \theta_1^{est}} [\theta_1 - \theta_1^{est}]^2 - K_2 \frac{T^2\bar{p}_2^2}{\theta_2 \theta_2^{est}} [\theta_2 - \theta_2^{est}]^2$$

Since $\theta_1$ and $\theta_2$ are positive parameters, and by selecting $K_i = -c_i sign(\theta_i^{est})$, namely, $\beta(x_k)$ as follows:

$$\beta(x_k) = \begin{bmatrix} -c_1 sign(\theta_1^{est}) & 0 \\ 0 & -c_2 sign(\theta_2^{est}) \end{bmatrix} \underbrace{\frac{A(x_k)^T}{\alpha[\bar{\sigma}(A(x_k)^T A(x_k)) + 1]}}_{\eta} \quad (4.38)$$

where $c_i > 0$ are arbitrary free positive scalars. Then, (4.37) becomes:

$$\frac{c_1 sign(\theta_1^{est})T^2\bar{p}_1^2}{\eta \, \theta_1 \theta_1^{est}}[\theta_1 - \theta_1^{est}]^2 + \frac{c_2 sign(\theta_2^{est})T^2\bar{p}_2^2}{\eta \, \theta_2 \theta_2^{est}}[\theta_2 - \theta_2^{est}]^2 \geq 0 \quad (4.39)$$

Hence, the first condition is satisfied.

Secondly, let us now consider the second condition:

$$\|(\theta_k^{est} - \theta)\| \geq \|\beta(x_k)A(x_k)[\phi(x_k,\theta) - \phi(x_k,\theta_k^{est})]\| \quad (4.40)$$

Again, replacing all the terms of (4.40) with their equivalents yields:

$$\|[\theta_k^{est} - \theta]\| \geq \left\| \begin{bmatrix} \frac{c_1 sign(\theta_1^{est})T^2\bar{p}_1^2}{\eta \, \theta_1 \theta_1^{est}} & 0 \\ 0 & \frac{c_2 sign(\theta_2^{est})T^2\bar{p}_2^2}{\eta \, \theta_2 \theta_2^{est}} \end{bmatrix} [\theta - \theta_k^{est}] \right\| \quad (4.41)$$

Noting that if the following inequality is satisfied, then inequality (4.41) is satisfied as well, which means that $\psi(x_k, \theta)$ is locally Lipschitz with $L < 1$.

$$\left\| \frac{1}{\alpha[\bar{\sigma}(A(x_k)^T A(x_k)) + 1]} \begin{bmatrix} \frac{c_1 sign(\theta_1^{est})T^2\bar{p}_1^2}{\theta_1 \theta_1^{est}} & 0 \\ 0 & \frac{c_2 sign(\theta_2^{est})T^2\bar{p}_2^2}{\theta_2 \theta_2^{est}} \end{bmatrix} \right\| < 1 \quad (4.42)$$

Thus, the above inequality is reduced to the following relation

$$\frac{\alpha}{c_i} > \frac{sign(\theta_i^{est})\, T^2\bar{p}_i^2}{[\bar{\sigma}(A(x_k)^T A(x_k)) + 1]\, \theta_i \theta_i^{est}} \quad (4.43)$$

Therefore, to satisfy the condition strictly, the estimator parameter can be selected as follows considering the lower bound of $\theta_i$ since it is always has a positive value where $\delta > 0$ is a positive free constant.

$$c_i = \frac{[\bar{\sigma}(A(x_k)^T A(x_k)) + 1]\, \theta_i \theta_i^{est}}{sign(\theta_i^{est})\, T^2\bar{p}_i^2(\delta + 1)} \alpha$$

Thus, now the proposed adaptive IDA-PBC controller in (3.28) can be straightforwardly constructed by making use of the derived non-linearly parameterized discrete-time gradient of the uncertain energy function of the system, the desired energy function and it's discrete-gradient, and the selected design function $\beta(x_k)$ above.

### 4.2.2 Simulation results

In the simulations, parameters of the considered desired energy function are taken as: $k_1 = 0.214$, $a_1 = 2$, $a_2 = -3$, and $a_3 = 5$. The damping injection control parameter is chosen as $K_v = 10$, and the estimator parameters are chosen as $c_1 = 6$, $c_2 = 2$, and $\alpha = 1$. The simulations are carried out under-sampling time $T_s = 0.01s$. The initial conditions are selected as $[q(0) \quad p(0)]^T = [2 \quad 0 \quad 0 \quad 0]^T$ and $[\theta_{est}(0)] = [0 \quad 0]^T$. The nominal value of the system parameter is taken as $m = 1kg$, $L = 1m$, $I_1 = 0.1kg.m^2$ and $I_2 = 0.1kg.m^2$. The desired positions as $q_1^* = q_2^* = 0$.

The simulation results for this example are given in Figs. 7-9. The closed-loop system dynamics $(q_1, q_2, p_1, p_2)$ for three different cases are illustrated together with desired system dynamics in Fig. 7. In the figure, the dynamics of the desired system are depicted in black color. The red color depicts the closed-loop system dynamics for the case where actual system parameters are different from the nominal values of the system parameters, namely $[I_1 \quad I_2]^T = [0.15 \quad 0.08]^T$, and the system is controlled by a non-adaptive controller such that actual system parameter values are not known by the non-adaptive controller, and thus the non-adaptive controller uses the nominal values of system parameters instead, namely, $[I_1 \quad I_2]^T = [0.1 \quad 0.1]^T$. The blue color depicts the closed-loop system dynamics for the case where the actual system parameters are $[I_1 \quad I_2]^T = [0.15 \quad 0.08]^T$ and the system is controlled by the proposed adaptive controller that uses the estimated parameters obtained by the proposed parameter estimator. Finally, the green color depicts the closed-loop system dynamics for the case where the system is controlled by the controller that knows and uses the actual system parameters, namely, $[I_1 \quad I_2]^T = [0.15 \quad 0.08]^T$. The figure illustrates that the adaptive controller preserves the closed-loop performance under uncertainty successfully while the non-adaptive controller fails to maintain the closed-loop performance for this example as well.

The dynamics of the estimated parameters are presented in Fig. 8. It can be seen in the figure that the estimated parameters converge to the true parameters successfully. Besides, the corresponding control signals for the presented three cases are shown in Fig. 9.

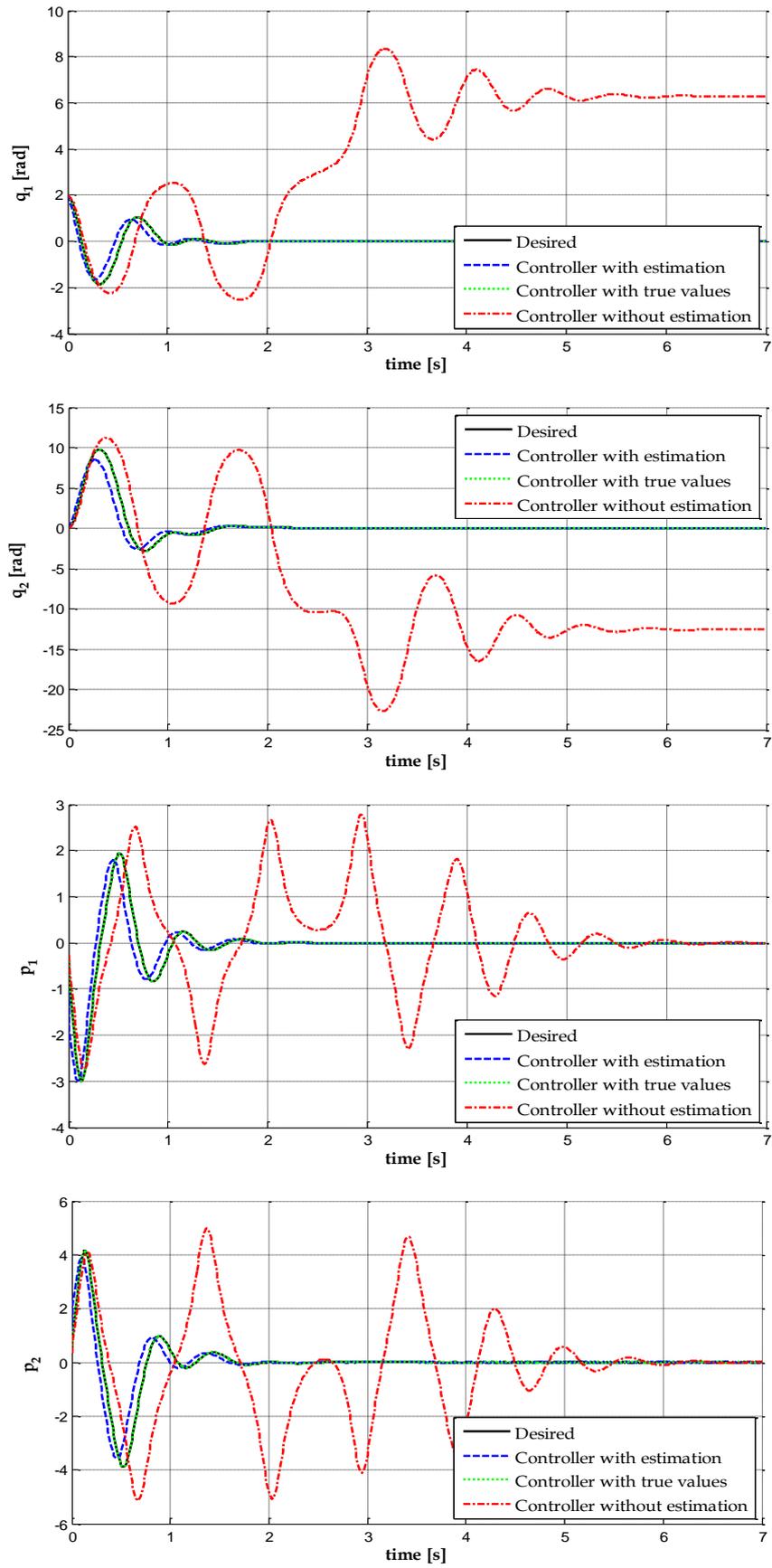

**FIGURE 7** The closed-loop dynamics under control of adaptive and non-adaptive controllers

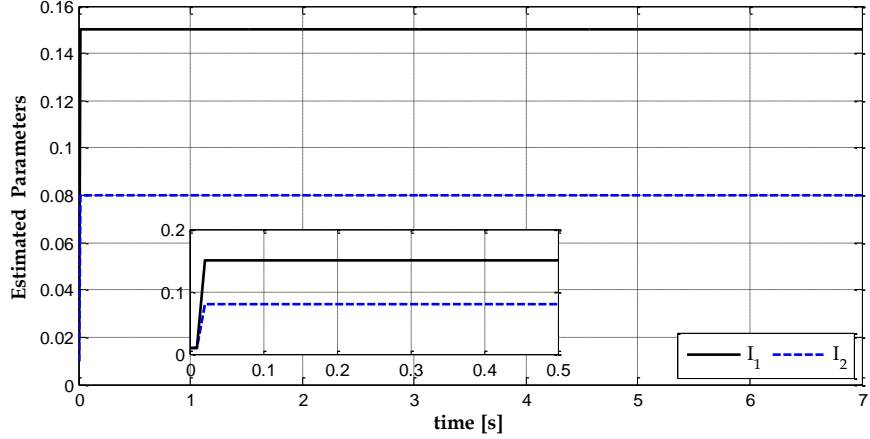

**FIGURE 8** The dynamics of the estimated parameters

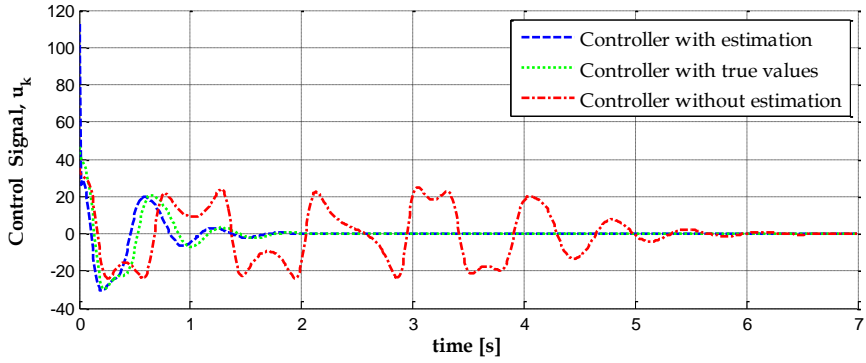

**FIGURE 9** The control signals for the different considered cases

## 5. CONCLUSIONS

A discrete-time I&I-based adaptive IDA-PBC control design for nonlinearly parameterized port-controlled Hamiltonian systems with parameter uncertainties in the energy function is presented in this paper. The conditions on the stability of the Immersion and Invariance estimator are derived. A proper structure is proposed for the free design function of the estimator such that it facilitates and includes some other free functions to construct estimators that satisfy the derived stability conditions. The free parameters exist in the proposed structure that enables tuning for the behavior of the estimator error system. The Lyapunov asymptotic stability of the estimator error dynamics and the local asymptotic stability of the closed-loop system dynamics obtained with the proposed adaptive IDA-PBC controller are proved for any sufficiently large set assuming the derived conditions are satisfied. The performance of the proposed adaptive control method is investigated on two physical systems via simulations. The obtained results verified the effectiveness of the proposed adaptive IDA-PBC controller in comparison to the non-adaptive IDA-PBC controller.

## CONFLICT OF INTEREST
The authors declare that they have no conflicts of interest.